\documentclass[preprint,12pt]{elsarticle}

\usepackage{amssymb}
\usepackage{wasysym}

\usepackage{epstopdf}
\usepackage{url}
\usepackage[usenames,dvipsnames]{color}
\usepackage{listings}

\journal{Journal of Computational Physics}

\begin{document}

\begin{frontmatter}


 \title{Title\tnoteref{label1}}

\title{The Fluid-Kinetic Particle-in-Cell Solver for Plasma Simulations}


\author{Stefano Markidis}
\address{High Performance Computing and Visualization (HPCViz) Department, KTH Royal Institute of Technology, Stockholm, Sweden}
\author{Pierre Henri}
\address{Universit\'e de Nice Sophia Antipolis, CNRS, Observatoire de la C\^ote d'Azur, Nice, France}
\author{Giovanni Lapenta}
\address{Centre for Plasma Astrophysics, Katholieke Universiteit Leuven, Celestijnenlaan 200B, B-3001 Leuven, Belgium}
\author{Kjell R\"onnmark}
\address{Physics Department, Ume\aa \ University, Ume\aa, Sweden}
\author{Maria Hamrin}
\address{Physics Department, Ume\aa \ University, Ume\aa, Sweden}
\author{Zakaria Meliani}
\address{LUTH - Astronomical Observatory of Paris, Paris, France}
\author{Erwin Laure}
\address{High Performance Computing and Visualization (HPCViz) Department, KTH Royal Institute of Technology, Stockholm, Sweden}

\begin{abstract}
A new method that solves concurrently the multi-fluid and Maxwell's equations has been developed for plasma simulations. By calculating the stress tensor in the multi-fluid momentum equation by means of computational particles moving in a self-consistent electromagnetic field, the kinetic effects are retained while solving the multi-fluid equations. The Maxwell's and multi-fluid equations are discretized implicitly in time enabling kinetic simulations over time scales typical of the fluid simulations. The  fluid-kinetic Particle-in-Cell solver has been implemented in a three-dimensional electromagnetic code, and tested against the ion cyclotron resonance and magnetic reconnection problems. The new method is a promising approach for coupling fluid and kinetic methods in a unified framework.  

\end{abstract}


\end{frontmatter}


\section{Introduction}
Simulations of plasmas are an important tool to study and understand the plasma behavior in fusion devices, in astrophysical and space environments. Plasmas are an archetype of multi-scale multi-physics problem, and simulations capable to encompass the whole range of scales in a plasma are very challenging. In fact, the separation of micro and macro scales in plasmas is due to the large difference of mass of electrons and protons (ions). Electrons are 1836 times lighter than protons and they are characterized by a fast dynamics. Instead, protons are heavier and their dynamics develops over larger spatial scales and slower time scales. The correct description of plasma micro scale phenomena such as the interaction of electrons and protons with waves, collisionless shocks and magnetic reconnection requires kinetic modeling. The kinetic approach solves an equation for the evolution of the distribution function of electrons and ions, and it is very computationally expensive. On the other hand, plasma can be treated macroscopically as a neutral fluid governed by the fluid and Maxwell's equations. The fluid modeling is a very convenient approach to describe the large scale and low frequency plasma dynamics but it neglects the micro physics in spatial regions smaller than ion gyro radius and in time periods shorter than ion gyro-period~\cite{hazeltine1998framework}. Because the kinetic modeling comes at huge computational cost, with the current computer resources it can be only applied to study plasma in small regions of space where kinetic effects are known to be important (i.e. magnetic reconnection sites, collisionless shock fronts, double layer regions)~\cite{lapenta2011space}. On the other hand, the fluid approach can describe the macroscopic behavior of plasma over very large scale with the current computational resources. For these reasons, it is desirable to have a computational framework where both fluid and kinetic descriptions are present. The fluid model can be used in computational regions where kinetic effects are not important, while the kinetic approach can be applied to small regions where kinetic effects are dominant~\cite{Sugiyama:2007, Ishiguro:2010}. An alternative approach consists of having both kinetic and fluid models present in the simulation and using kinetic modeling to provide a closure to the fluid equations. This work follows this alternative approach with the goal of achieving a description of plasma within a unified framework comprising both fluid and kinetic modeling.  \\

Many computational studies have been made in the last 15 years on the coupling fluid and kinetic approaches. For instance, coupling of Boltzmann equation and Euler or Navier-Stokes equations~\cite{Bourgat:1996, Degond:2005, Degond:2010} have been developed to solve the atmospheric entry problem in aerodynamics. The equivalent problem for the plasma dynamics is the coupling between the Vlasov (the collisionless Boltzmann equation) and fluid equations. In this article, we present the simultaneous computation of two coupled models, one describing the evolution of the large scale structures with a fluid model for different plasma species (electron, ions), the other describing the evolution of structures at small scales with a kinetic modeling solving the Vlasov equation with the Particle-in-Cell (PIC) technique~\cite{birdsall, hockney}. The model described in this paper differs from previous fluid-PIC coupling methods~\cite{Sugiyama:2007, Ishiguro:2010, SWIFF:2013}, where kinetic and fluid models are solved in separate regions of space and coupled at the interface of the simulation areas. In the kinetic-fluid solver, fluid and kinetic equations are solved self-consistently and simultaneously in a unified framework.   \\

We propose a new fluid-kinetic PIC method. This solver belongs to the category of the implicit PIC methods. In fully implicit PIC methods ~\cite{Markidis:2011,Chacon:2011}, the Maxwell's equations and particle equations of motion are computed concurrently with a non linear solver until convergence is achieved. One of the main advantages of these methods is that they are energy conserving and provide a correct description of particle acceleration~\cite{Lapenta:2011}. Instead, in the semi-implicit PIC methods, fields and particle equations of motion are decoupled and solved in separate parts of the computational cycle (mover and Maxwell solver). The field equations for the electromagnetic field are modified and they include approximate current and charge densities and a numerical implicit susceptibility. In the implicit direct method~\cite{langdon1983direct,hewett1987electromagnetic,drouin2010particle}, these equations are calculated by computing orbits of particles, while in the implicit moment method the fluid equations~\cite{Mason,Brackbill:1982,Ricci2002} are used. The fluid-kinetic PIC solver is similar to the implicit moment PIC method. In both methods, the fluid equations are used to evaluate the electromagnetic field. However, fluid equations are explicitly included in the fluid-kinetic PIC solver. Instead, they are not in the implicit moment method and they are used to give an estimate of the charge and current densities as sources and as implicit susceptibility in the Maxwell's equations. 

The fluid-kinetic PIC method removes the approximations introduced by the implicit moment method. In addition, the fluid-kinetic PIC method extends and improves the particle-fluid model proposed by Vedin and R\"onnmark in Refs.~\cite{Vedin:2006, Vedin:2007} that has been successfully used for the simulation of auroral electrons. Differently from the particle-fluid model, the fluid-kinetic PIC method is formulated for three dimensional geometry and with no assumption on the magnetic field strength.   \\

This article is organized as follows. Section 2 presents the governing equations, and Section 3 describes their discretization in time and space. Section 4 explains the fluid-kinetic PIC algorithm and its implementation in a three dimensional electromagnetic code. Simulation results obtained with the fluid-kinetic PIC solver, and a performance analysis of the new method are presented respectively in Sections 5 and 6. Finally, Section 7 summarizes the fluid-kinetic PIC method, emphasizes its advantages, and outlines future work on the fluid-kinetic PIC method.

\section{Governing Equations}
The proposed fluid-kinetic solver is a PIC method~\cite{birdsall,hockney}. The PIC method solves the Vlasov equation by sampling the initial distribution function with computational particles and following their trajectories in time by solving the particle equation of motion~\cite{birdsall,hockney}. At each PIC computational cycle, the distribution function can be reconstructed by using the computational particles. The electric and magnetic fields, that act on the particles, and their sources (charge and current densities) are defined on a grid. The link between particles and quantities defined on the grid is provided by the interpolation functions that allow to calculate first the charge and current densities from interpolating particle positions and velocities, and then it allows to calculate the electric and magnetic fields acting on the particles from the values of the electromagnetic field defined on the grid points.  \\

The governing equations for the fluid-kinetic PIC solver are the multi-fluid equations, Maxwell's equations and particle equations of motion. The multi-fluid equations and Maxwell's equations are solved on the grid, while the equation of motion are solved for each computational particle. \\

The fluid continuity and momentum equations for the species $s$ are (here and thereafter in CGS units):
 \begin{equation}
 \label{Fluid}
\left\{
\begin{array}{l}
\partial \rho_s/\partial t  +  \nabla \cdot \mathbf{J}_s = 0 \\
\partial \mathbf{J}_s/\partial t  = (q/m)_s (\rho_s \mathbf{E} + (\mathbf{J}_s \times \mathbf{B})/c -\nabla \mathcal{T}_s)\\
\end{array}
\right.
\end{equation}
where $\rho_s$ and $\mathbf{J}_s$  are respectively the charge and current density for the species $s$. Species can represent electron, and ions of different kinds (protons, Helium nuclei,...). $\mathcal{T}_s = \mathbf{J}_s \mathbf{J}_s/\rho_s + p_s$ and $p_s$ are respectively the stress and pressure tensors. $\mathbf{E}$ and $\mathbf{B}$ are the electric and magnetic fields, and $c$ is the speed of light in vacuum. In this work, we do not include an equation for energy conservation, but we provide a closure equation for calculating the stress tensor with computational particles. In the next subsection, a technique to evaluate the stress tensor by computational particles is described in detail.   \\

Together with the multi-fluid equations, the Maxwell's equations are solved. The evolution of the electric and magnetic fields is determined by solving the Maxwell's equations:
 \begin{equation}
 \label{Maxwell}
\left\{
\begin{array}{l}
\nabla \cdot \mathbf{E} =  4 \pi  \rho \\
\nabla \cdot \mathbf{B} = 0 \\
 {1}/{c}\,{\partial \mathbf{E}}/{\partial t}  = \nabla \times \mathbf{B}  - {4 \pi}/{c} \mathbf{J}  \\
 {1}/{c}\,{\partial \mathbf{B}}/{\partial t}  = - \nabla \times \mathbf{E},
\end{array}
\right.
\end{equation}

The fluid and Maxwell's equations are coupled as the charge an current densities depend on the electric  and magnetic field (Eq.~\ref{Fluid}) and vice-versa the electromagnetic field depends on charge an current densities (Eq.~\ref{Maxwell}).

\subsection{Particle Closure Equation}
Different closure equations can be provided to calculate the stress tensor $\mathcal{T}$ in Eq.~\ref{Fluid}~\cite{hazeltine1998framework}. In this work, we intend to retain the kinetic effects by calculating the stress tensor using the computational particles of the PIC method. Each computational particle is characterized by a position $\mathbf{x}_p$ and a velocity $\mathbf{v}_p$, whose evolution is described by the equation of motion:
\begin{equation}
\label{motion}
\left\{
\begin{array}{l}
{d {\bf x}_p}/{dt} ={\bf v}_p \\
{d {\bf v}_p}/{dt} =q_s/m_s\left( {\bf E}_p + {\bf v}_p/c \times {\bf B}_p\right),
\end{array}
\right.
\end{equation}
where $q_s/m_s$ are the charge to mass ratio of the species $s$. ${\bf E}_p $, and ${\bf B}_p $ are the electric and magnetic fields acting on the particle $p$ and they are calculated by interpolation from ${\bf E}_g$ and ${\bf B}_g $, the values of the electric and magnetic field on the $N_g$ grid points, through the use of the interpolation function $W({\bf x}_g-{\bf x}_p) $:
\begin{equation}
{\bf E}_p=\sum_g^{N_g} {\bf E}_g W({\bf x}_g-{\bf x}_p) \quad \quad {\bf B}_p=\sum_g^{N_g} {\bf B}_g W({\bf x}_g-{\bf x}_p) .
\label{interp}
\end{equation}
Typically, the Cloud-in-Cell interpolation functions~\cite{birdsall,hockney} are used:
\begin{equation}
 \label{interpW}
W({\bf x}_g-{\bf x}_p) =
\left\{
\begin{array}{l}
 1 - |{\bf x}_g-{\bf x}_p|/\Delta x \quad \textup{if} \quad |{\bf x}_g-{\bf x}_p| < \Delta x \\
0  \quad \textup{otherwise}  .
\end{array}
\right.
\end{equation}
In the PIC method the charge and current densities ($\rho,\mathbf{J}$) defined on the grid are calculated with interpolation functions:
\begin{equation}
\{ \rho_s,\mathbf{J}_s\}_g  =  \sum_p^{N_p} q_s \{1,  \mathbf{v}_p \} W({\bf x}_g-{\bf x}_p) .
\end{equation}
In the fluid-kinetic PIC method, the interpolation functions are used to calculate the stress tensor $\mathcal{T}_s$ also. The closure equation for the stress tensor is provided by particles in this way:
\begin{equation}
(\mathcal{T}_s)_g  =  \sum_p^{N_p} q_s \mathbf{v}_p  \mathbf{v}_p W({\bf x}_g-{\bf x}_p) .
\end{equation}
The idea of using computational particles to provide a closure equation is not new, but it has been previously introduced in semi-implicit PIC methods~\cite{brackbill-cohen-85, Brackbill:1982} and in the fluid-particle model~\cite{Vedin:2006,Vedin:2007}.

\section{Discretization of Governing Equations}
The fluid continuity and momentum equations together with the second order formulation of MaxwellÕs equations are solved concurrently using the finite difference box scheme~\cite{Morton}. We present the discretization in time and in space and a discussion of the numerical stability of the fluid-kinetic PIC method in the following subsections.

\subsection{Time Discretization}
Both fluid and Maxwell's equations (Eqs. \ref{Fluid}, \ref{Maxwell}) are discretized implicitly in time and solved concurrently in the fluid-Maxwell solver. The fluid equations are discretized in time as follows:
 \begin{equation}
 \label{FluidDiscretized}
\left\{
\begin{array}{l}
  {\rho_s^{n+1}  -  \rho_s^{n}} +  {\Delta t}  \nabla \cdot \mathbf{J}_s^{n+1/2} = 0 \\
  {\mathbf{J}_s^{n+1} - \mathbf{J}_s^{n}} = {\Delta t}  (q/m)_s (\rho_s^{n}  \mathbf{E}^{n+\theta} + (\mathbf{J}_s^{n+1/2} \times \mathbf{B}^{n})/c -\nabla \mathcal{T}_s^{n})\\
\end{array}
\right.
\end{equation}
where $n+1$ and  $n$ are the time levels, and $\theta$ is the de-centering parameter that can be chosen between 0.5 and 1.0. The quantity $\mathbf{E}^{n+{\theta}}$ is defined as:
\begin{equation}
\mathbf{E}^{n+{\theta}} = \theta \mathbf{E}^{n+{1}} + (1- {\theta})\mathbf{E}^{n}
\end{equation}
The term $\mathbf{J}_s^{n+1/2}$ is defined as time average $(\mathbf{J}_s^{n+1} + \mathbf{J}_s^{n})/2$.
The Ampere's and Faraday's laws are discretized implicitly in time:
\begin{equation}
 \label{MaxwellDiscretized}
\left\{
\begin{array}{l}
c \theta \Delta t \nabla \times \mathbf{E^{n+\theta}} + \mathbf{B}^{n+1} - \mathbf{B}^{n} =  0, \\
c \theta \Delta t \nabla \times \mathbf{B^{n+\theta}} - ({\mathbf{E}^{n+1} - \mathbf{E}^{n}})  =   4 \pi  \theta \Delta t \mathbf{J}^{n+1/2}  \\
\end{array} .
\right.
\end{equation}

Taking the curl of Faraday's law allows to eliminate $\mathbf{B^{n+\theta}}$ and recover an expression for $\mathbf{E}^{n+\theta}$:

 \begin{equation}
 \label{Maxwell3}
 (c \theta \Delta t)^2 \nabla  \times  \nabla  \times   \mathbf{E^{n+\theta}}  +  \mathbf{E^{n + 1}} -   \mathbf{E^{n
 }}= (\theta \Delta t)(c \nabla  \times \mathbf{B}^{n} - 4 \pi\mathbf{J}^{n+1/2} )
 \end{equation}
Using the vector identity  $\nabla  \times  \nabla  = \nabla \nabla \cdot - \nabla^2$, and the Gauss' law $\nabla  \cdot  \mathbf{E}^{n+\theta} = 4 \pi\rho^{n+1/2} = 4 \pi(\rho^{n+1} + \rho^{n})/2 $, we obtain:
 \begin{equation}
 \label{Maxwell4}
 \mathbf{E^{n+\theta}} - (c \theta \Delta t)^2 \nabla^2  \mathbf{E^{n+\theta}} =  \mathbf{E^{n}}    + c \theta \Delta t  (\nabla  \times \mathbf{B}^{n} - 4 \pi (\mathbf{J}^{n+1/2}  + c \theta \Delta t \nabla \rho^{n+1/2}  )) .
 \end{equation}
 This equation is solved to calculate $\mathbf{E^{n+\theta}}$.   \\

In summary, the unknowns of the problem are $\mathbf{E^{n+\theta}}$, $\rho_s^{n+1}$, and $\mathbf{J}_s^{n+1}$ and the following equations are solved concurrently on the grid in the fluid-Maxwell solver in the kinetic-fluid PIC method:
 \begin{equation}
 \label{kinflusolv}
\left\{
\begin{array}{l}
  {\rho_s^{n+1}  -  \rho_s^{n}}  +  {\Delta t} \nabla \cdot \mathbf{J}_s^{n+1/2} = 0 \\
  {\mathbf{J}_s^{n+1} - \mathbf{J}_s^{n}}  = (q/m)_s {\Delta t} (\rho_s^{n}  \mathbf{E}^{n+\theta} + (\mathbf{J}_s^{n+1/2} \times \mathbf{B}^{n})/c -\nabla \mathcal{T}_s^{n})\\
 \mathbf{E^{n+\theta}} - (c \theta \Delta t)^2 \nabla^2  \mathbf{E^{n+\theta}} =  \mathbf{E^{n}}    + c \theta \Delta t  (\nabla  \times \mathbf{B}^{n} - 4 \pi (\mathbf{J}^{n+1/2}  + c \theta \Delta t \nabla \rho^{n+1/2}  ))  \\
\end{array}
\right.
\end{equation}
Once the electric field $ \mathbf{E^{n+\theta}}$ is calculated, the magnetic field is advanced in time by solving the discretized FaradayÕs law:
\begin{equation}
\frac{ \mathbf{B}^{n+{1}} - \mathbf{B}^{n} }{\Delta t}= - \nabla \times \mathbf{E}^{n+\theta} .
\end{equation}
When $\theta=0.5$ the fluid-Maxwell solver method results second order accurate in time. \\

In the fluid-kinetic PIC method, the stress tensor $\mathcal{T}_s^{n}$ is calculated from the particles positions and velocities with the interpolation functions. Eqs. \ref{motion} are differenced in time using the implicit midpoint integration rule~\cite{Hairer:2002}:
\begin{equation}
\label{dif_eom}
\left\{
\begin{array}{l}
\mathbf{v}_p^{n+1} = \mathbf{v}_p^{n} + q_s /m_s \Delta t (\mathbf{\bar{E}}_p +  \mathbf{\bar{v}}_p/c \times \mathbf{\bar{B}}_p)  \\
\mathbf{x}_p^{n+1} = \mathbf{x}_p^{n} + \mathbf{\bar{v}}_p\Delta t
\end{array} 
\right. .
\end{equation}
It is possible to rewrite Equations \ref{dif_eom} in terms of $\mathbf{\bar{v}}_p$ after a series of algebraic manipulations~\citep{markidisPHDthesis}:
\begin{eqnarray}
\label{vhat2}
\tilde{\mathbf{v}}_p&=&\mathbf{v}_p^n+\frac{q_s\Delta t}{2m_s}\mathbf{\bar{E}}_p\\
\label{vn+1/2}
\mathbf{\bar{v}}_p&=&\frac{\tilde{\mathbf{v}}_p+\frac{q_s\Delta
t}{2m_s c}\bigl(\tilde{\mathbf{v}}_p\times\mathbf{\bar{B}}_p+\frac{q_s\Delta
t}{2m_s c}(\tilde{\mathbf{v}}_p\cdot\mathbf{\bar{B}}_p)\mathbf{\bar{B}}_p\bigr)}{(1+\frac{q_s^2\Delta t^2}{4m_s^2c^2}{\bar{B}_p}^2)},
\end{eqnarray}
and the equation of motion for each particle becomes:
\begin{equation}
\label{dif_eom2}
\left\{
\begin{array}{l}
\mathbf{v}_p^{n+1} = 2 \mathbf{\bar{v}}_p - \mathbf{v}_p^{n}  \\
\mathbf{x}_p^{n+1} = \mathbf{x}_p^{n} + \mathbf{\bar{v}}_p\Delta t .
\end{array} 
\right. 
\end{equation}
This time discretization of the particle equation of motion is second order accurate in time~\cite{markidisPHDthesis}.

\subsection{Spatial  Discretization}
In this work, the multi-fluid and MaxwellÕs equations are differenced in space on a uniform Cartesian grid. The electric field, current densities, and stress tensor are defined on the nodes, while the magnetic field and the charge density are defined on the center cells. If the quantity $u_{i,j,k}$ is defined on vertices,
its spatial derivative $\partial u_{i,j,k}/\partial x$  is calculated at centers cell with indices $i+1/2, j+1/2, k+1/2$ as:
\begin{equation}
\frac{\partial u}{ \partial x}_{i+1/2, j+1/2, k+1/2} =   \frac{u_{i+1,j+1/2,k+1/2} - u_{i,j+1/2,k+1/2}}{\Delta x} .
\end{equation}
The values defined on centers are defined on centers by averaging the values defined on the nodes
\begin{equation}
u_{i,j+1/2,k+1/2} = \frac{1}{4}({u_{i,j,k} + u_{i,j+1,k+1} + u_{i,j+1,k}  + u_{i,j,k+1}}).
\end{equation}
The Laplacian operator is computed by combining the divergence and gradient operators. 
When spatial derivatives are approximated in this way, the system in Eq.~\ref{kinflusolv} results in a non-symmetric matrix.

\subsection{Stability Analysis}
A stability analysis of the fluid-kinetic PIC method for the electrostatic limit can be carried out in the same way of the implicit moment PIC method reported in Refs.~\cite{brackbill-cohen-85, Brackbill:1982} since the two PIC methods share the same governing equations. It results that in the fluid-kinetic PIC method the time step is not  constrained by the explicit PIC stability condition $\omega_{p}\Delta t < 2$ (where $\omega_{p}$ is the plasma frequency). We prove this experimentally in Section 6 reporting simulations with $\omega_{p}\Delta t >  2$. In addition, the fluid-kinetic PIC method is not subject to the Courant limit imposed by the propagation of light waves. Implicit PIC methods introduce selective damping and spectral compression of the unresolved wave modes~\cite{brackbill-cohen-85}. The selective numerical damping leads to a slight decrease of the total energy of the system and depends on the de-centering parameter $\theta$ and on the time step value $\Delta t$~\cite{brackbill-cohen-85, Brackbill:1982}. The fluid-kinetic PIC method removes the accuracy limit the implicit moment methods $v_{the} \Delta t/\Delta x$ that arises from using the Taylor expansion of the interpolation functions used in implicit moment method~\cite{lapenta2006kinetic}. As in other implicit PIC methods, the time step constraint is a Courant condition imposed by the propagation of acoustic waves~\cite{Mason,Brackbill:1982,Ricci2002}.

\section{Algorithm and Implementation}
\begin{figure}[ht]
\includegraphics[width=1 \columnwidth]{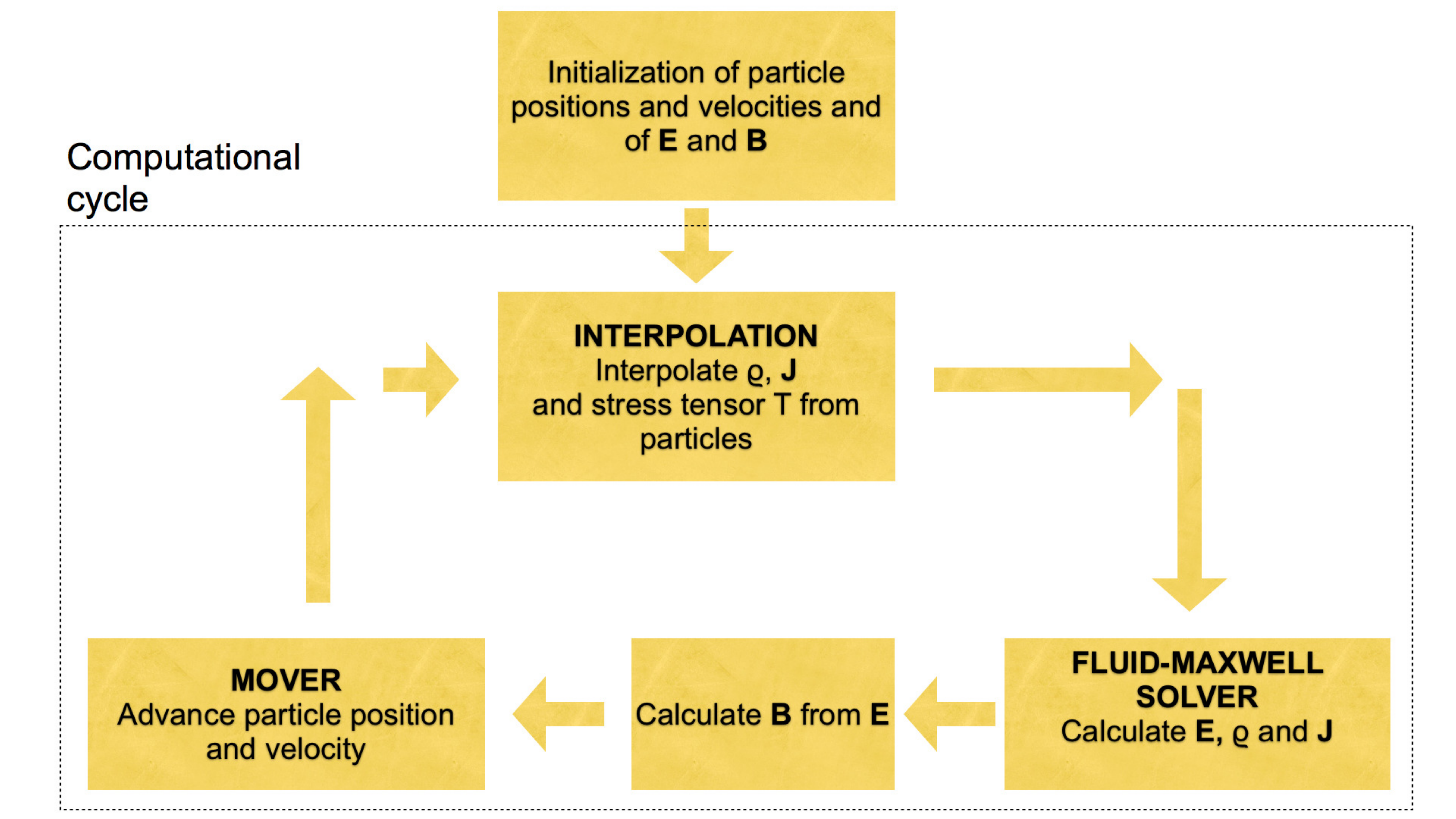}
\caption{Algorithm of the fluid-kinetic PIC method. After the simulation is initialized, a computational cycle is repeated. At each computational cycle, the charge and current densities, and the stress tensor are interpolated from particle positions and velocities, the fluid-Maxwell solver is executed, and particles are advanced in time.}
\label{Algorithm}
\end{figure}

The fluid-kinetic PIC solver algorithm is represented in Figure~\ref{Algorithm}. Initially, the system is set-up defining the electric and magnetic fields on the grid and sampling particle positions and velocities. A computational cycle is then repeated advancing in time the system variables. At each computational step, the charge and current densities and the stress tensor are computed by interpolation using particle positions and velocities. This step defines all the interpolated quantities at time $n$. Then the fluid-Maxwell solver computes the new values of the electric field and of the charge and current densities. The new magnetic field is calculated from the electric field. Finally, particles are advanced using the new values of the electric and magnetic fields.

\begin{figure}[ht]
\includegraphics[width=1 \columnwidth]{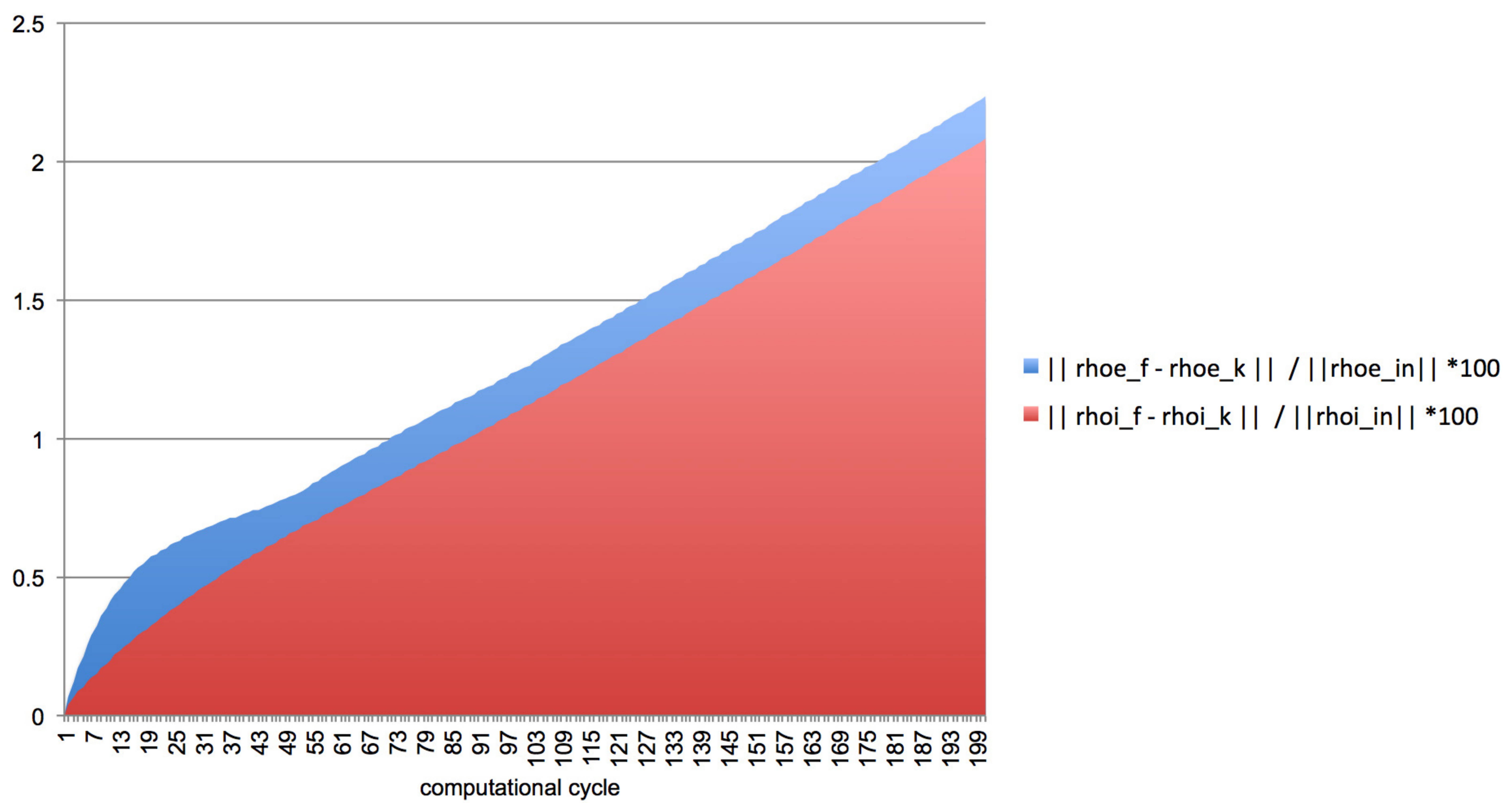}
\caption{Evolution of the norm of the difference of the fluid and kinetic charge densities normalized to norm of the initial charge density for electrons in blue and ions in red. The growing inconsistencies between fluid and kinetic densities make the PIC simulation unstable.}
\label{RHOKFdiff}
\end{figure}

\begin{figure}[ht]
\includegraphics[width=1 \columnwidth]{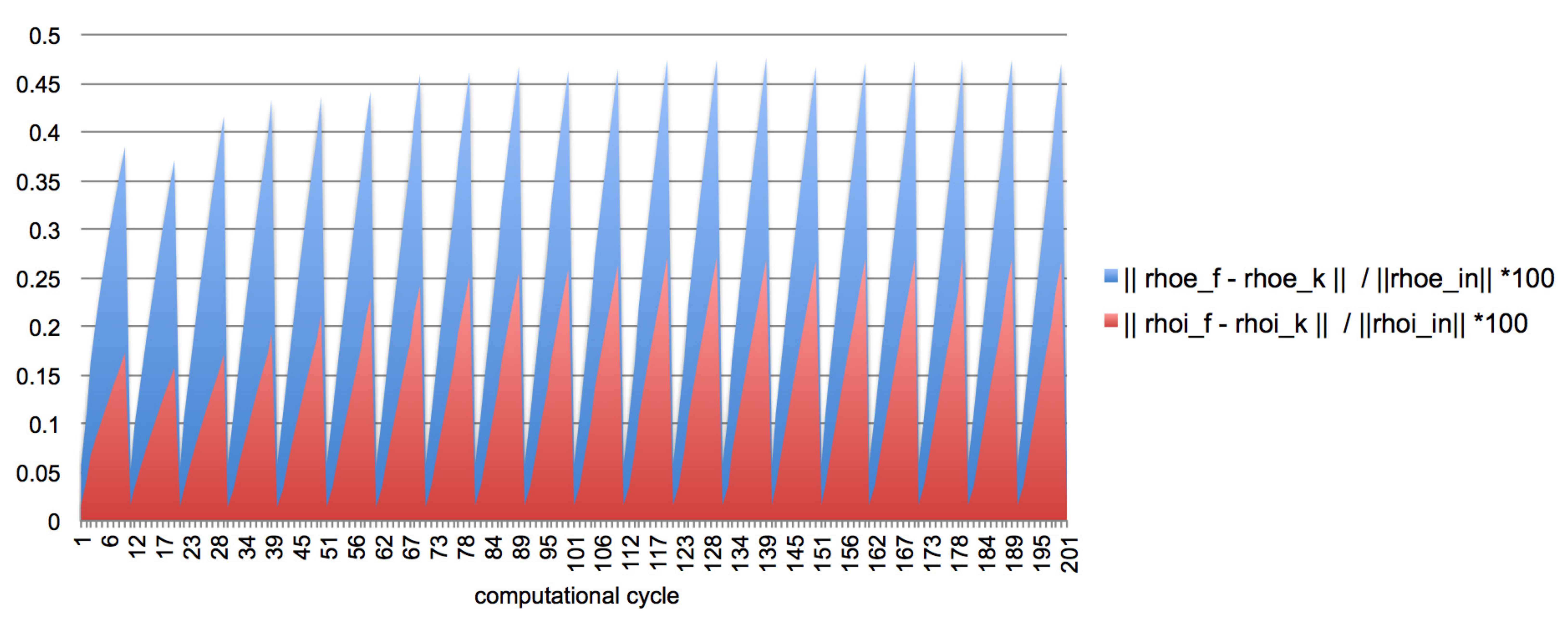}
\caption{Evolution of the norm of the difference of the fluid and kinetic charge densities normalized to norm of the initial charge density for electrons in blue and ions in red. In this simulation, the fluid charge and current densities are enforced to be equal to kinetic densities every ten computational cycles. }
\label{Resync}
\end{figure}

At a first analysis of the computational cycle, it might seem that there is no need for the current and charge density interpolation stage. In fact, these quantities are calculated in the fluid-Maxwell solver and they can be reused. However, the fluid and kinetic charge and current densities diverge in time. In Figure~\ref{RHOKFdiff}, the norm of the difference of the fluid and kinetic charge densities normalized to norm of the initial charge density during a magnetic reconnection simulation is shown in blue for electrons and in red for ions. During the evolution of the simulation, clearly the difference between the density and current of the particles and the density and current of the fluid increaseas. By using $\rho$ and $\mathbf{J}$ from fluid-Maxwell solver, the PIC simulation accumulates numerical errors and the growing inconsistencies between particle and fluid densities make the PIC simulation unstable.
For this reason, a correction is needed to keep the charge and current densities calculated by interpolation from particles (kinetic densities) equal to the charge and current densities from the fluid-Maxwell solver (fluid densities). One approach is to enforce fluid densities to be equal to the kinetic densities at each cycle (like in the fluid-kinetic PIC method) or at a given computational cycle. Figure~\ref{Resync} shows the norm of the difference of the fluid and kinetic charge densities normalized to norm of the initial charge density for a simulation where fluid and kinetic densities are enforced to be the same every ten computational cycles. It is clear that the difference between kinetic and fluid densities increases in time until the fluid densities are enforced to be the equal to the kinetic densities.

A second technique to enforce equality between fluid and kinetic quantities is proposed in Refs.~\cite{Vedin:2006,Vedin:2007}. In this alternative approach, the electric field acting on the particle $\mathbf{E}_p$ is corrected to take in account the particle density on timescales of the order of the inverse plasma frequency and on length scales similar to the Debye length. This correction is derived from the AmpereÕs and GaussÕ law~\cite{Vedin:2006,Vedin:2007}. \\

An additional correction needs to be applied to ensure that the charge density continuity equation is satisfied. This step is called "divergence cleaning". In the fluid-kinetic PIC code we use the technique proposed by Boris~\cite{boris1970relativistic}. Once the new electric Þeld $\mathbf{E}^{n+1}$ is calculated from Eq.~\ref{kinflusolv}, a correction $\delta \mathbf{E}$ for $\mathbf{E}^{n+1}$ is calculated as follow:

\begin{equation}
\label{Poisson}
 \delta \mathbf{E} = - \nabla \Phi \quad  \nabla^2 \Phi = \nabla \cdot \mathbf{E}^{n+1} - 4 \pi \rho .
\end{equation}

The discretized Eqs.~\ref{kinflusolv} and their boundary conditions form a non-symmetric linear system that is solved using the Generalized Minimal Residual (GMRes) method~\cite{saad2003iterative}. For solving Eq.~\ref{Poisson}, the Conjugate Gradient (CG) method~\cite{saad2003iterative} is used since the discretized equation leads to a symmetric matrix. The fluid-kinetic PIC code solves the particle equation of motion (Eqs. (~\ref{vn+1/2},~\ref{dif_eom2})) by an iterative method based on a fixed number of predictor-corrector iterations.  \\

A three-dimensional fluid-kinetic PIC solver is implemented starting from the iPIC3D code~\cite{iPIC}, a massively parallel fully kinetic code. The code is written in C++ and MPI. On multiprocessor architectures, the domain decomposition technique is used to divide the computational workload among processors. Particles are divided among processors also depending on their location, and communicated to adjacent processors if exiting from the processor domain. The parallelization of the code is based on MPI libraries and blocking parallel communication has been chosen for the communication among computing units.

\section{Results}
The new fluid-kinetic PIC method has been tested against a series of benchmark problems. In this section, the test of ion cyclotron resonance in two dimensional geometry and magnetic reconnection in two and three dimensional geometries are presented.

\subsection{Ion Cyclotron Resonance}
The first test simulates a plasma uniformly distributed in space and with a Maxwellian distribution in the velocities. The plasma consists of electrons and ions with $0.08 \ c$ and $0.02 \ c$ thermal velocities. The electron charge to mass ratio is 64. 250 computational particles per cell are used. A background magnetic field in the $x$ direction has been set to 0.02. The grid consists of $64\times64$ grid cells. The simulation box is $L_x \times L_y = 20 \ d_i \times 20 \ d_i$, where $d_i$ is the ion skin depth $d_i = c/\omega_{pi}$, where $\omega_{pi}$ is the ion plasma frequency. The time step is $2.4 / \omega_{p}$, $\theta = 1.0$, and 10,000 computational cycles are computed reaching a final simulation time $\Omega_{ci}t_{fin} = 60$, where $\Omega_{ci}$ is the ion cyclo-frequency. \\

\begin{figure}[ht]
\includegraphics[width=1 \columnwidth]{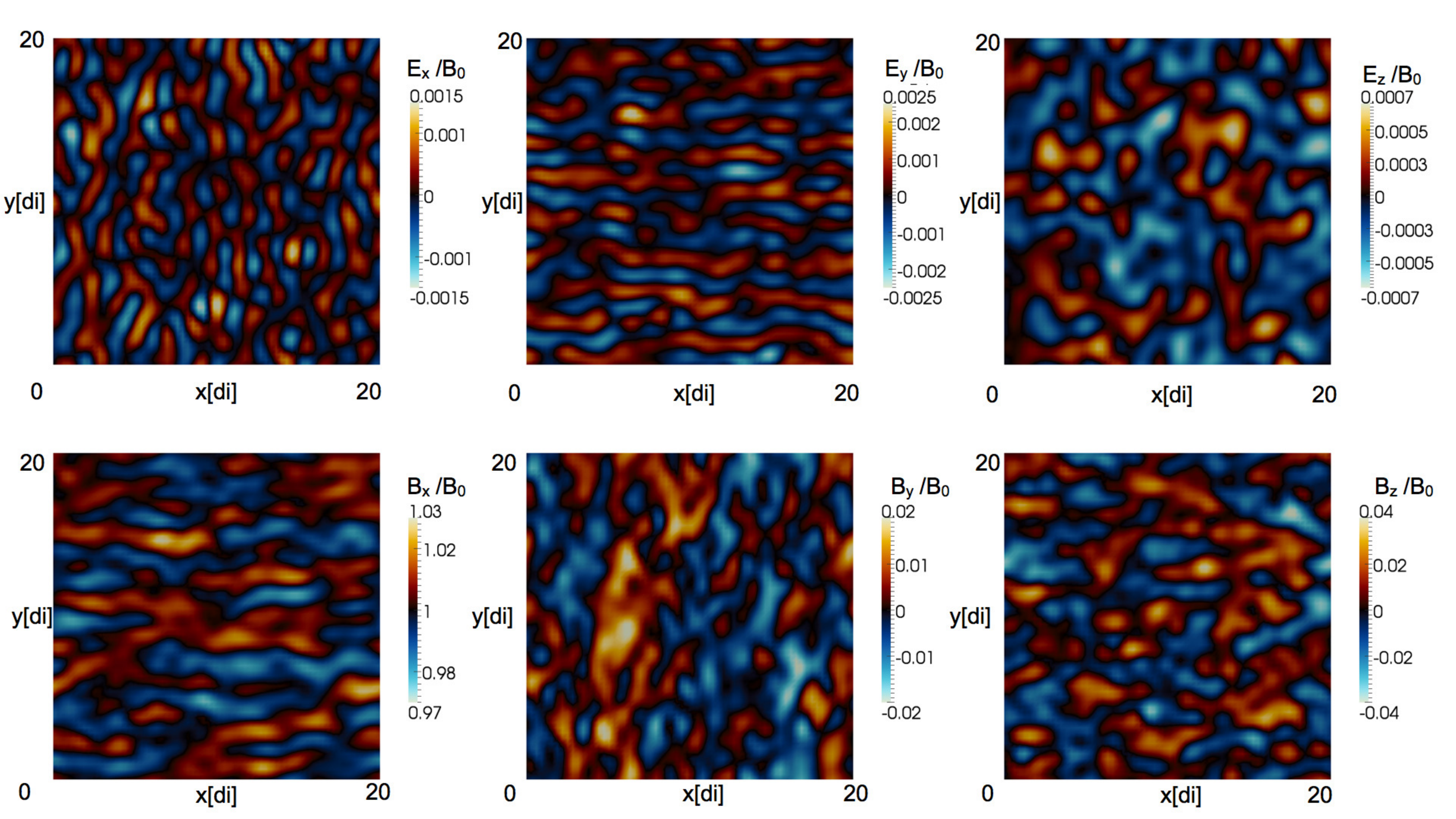}
\caption{Contour plot of the different components of the electric and magnetic fields at time $\Omega_{ci} \  t = 60$. The simulation shows the occurrence of different waves in the initially Maxwellian plasma.}
\label{PierreFig1}
\end{figure}

\begin{figure}[ht]
\includegraphics[width=1 \columnwidth]{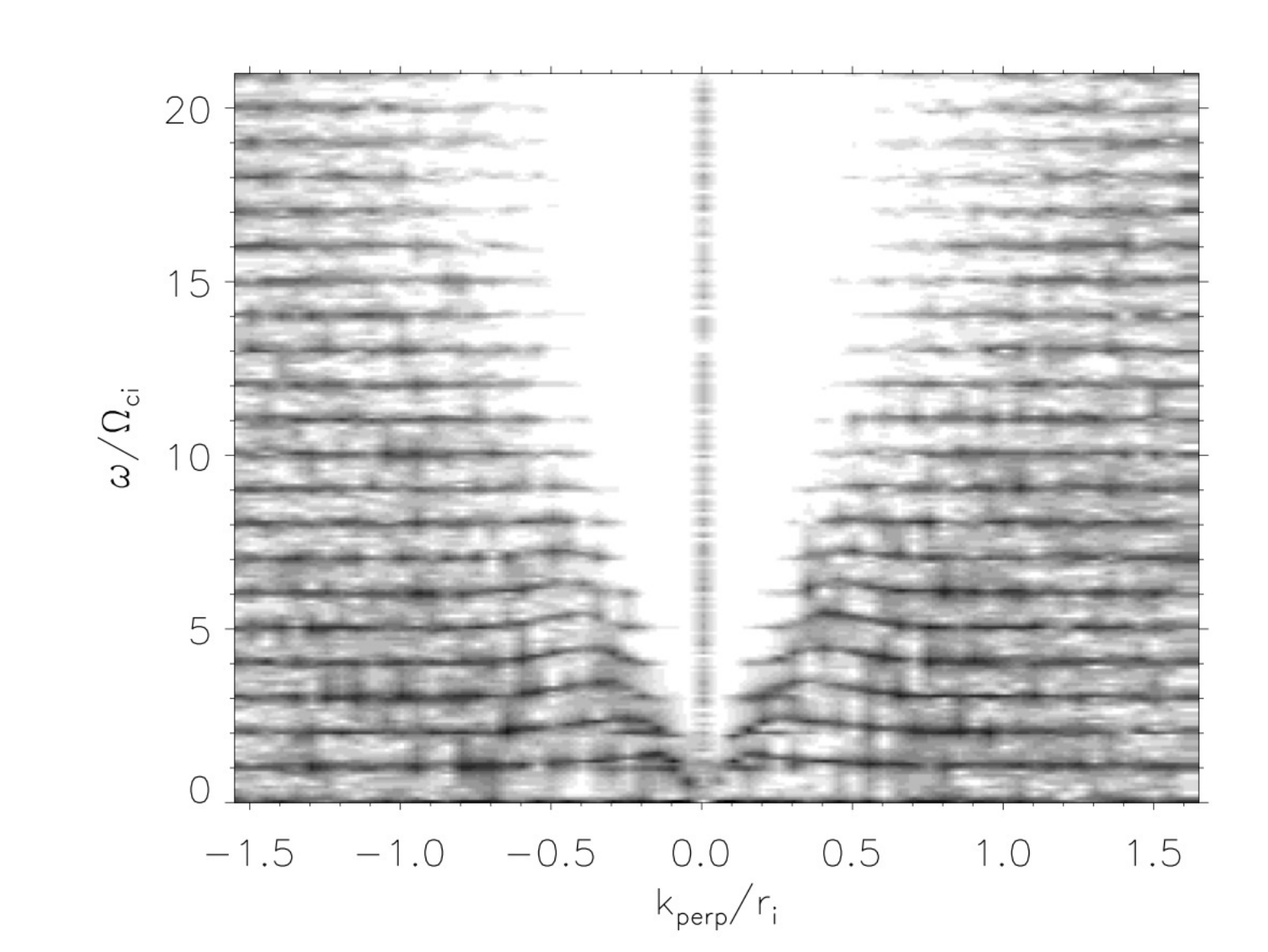}
\caption{Two-dimensional spectrum of the ion density field. The frequency $omega$ is normalized to the ion cyclotron frequency $\Omega_{ci}$ to ease the reading. The perpendicular wavevector $k_{perp}$ is normalized to the ion gyro-radius $r_i$. }
\label{fig:ionBernsteinRelDisp}
\end{figure}

The random noise in the particle velocity distribution is redistributed among the normal modes of the system. With the temporal and spatial parameters chosen for this simulation, we focus on the perturbations with a frequency of the order or larger than the ion cyclotron frequency $\Omega_{ci}$. The evolution of the system leads to the occurrence of different waves visible in the contour-plots of different components of the electromagnetic field in Figure \ref{PierreFig1}. To identify the wave modes, a spectral analysis of the simulation is carried out. Figure~\ref{fig:ionBernsteinRelDisp} shows the two-dimensional spectrum of the ion density, computed as the two-dimensional Fourier transform of the time-evolving ion density, averaged in the direction parallel to the magnetic field. It is expressed in the frequency - perpendicular wavevector plane. This spectrum enables to isolate the compressible modes propagating perpendicularly to the ambient magnetic field. It clearly shows the dispersion relations of the different ion Bernstein modes at the harmonics of the ion cyclotron frequency~\cite{hazeltine1998framework}. With this example we show that the fluid-kinetic PIC solver perfectly captures the cyclotron resonances, that are not captured by standard multi-fluid solvers. Ê

\begin{figure}[ht]
\includegraphics[width=1 \columnwidth]{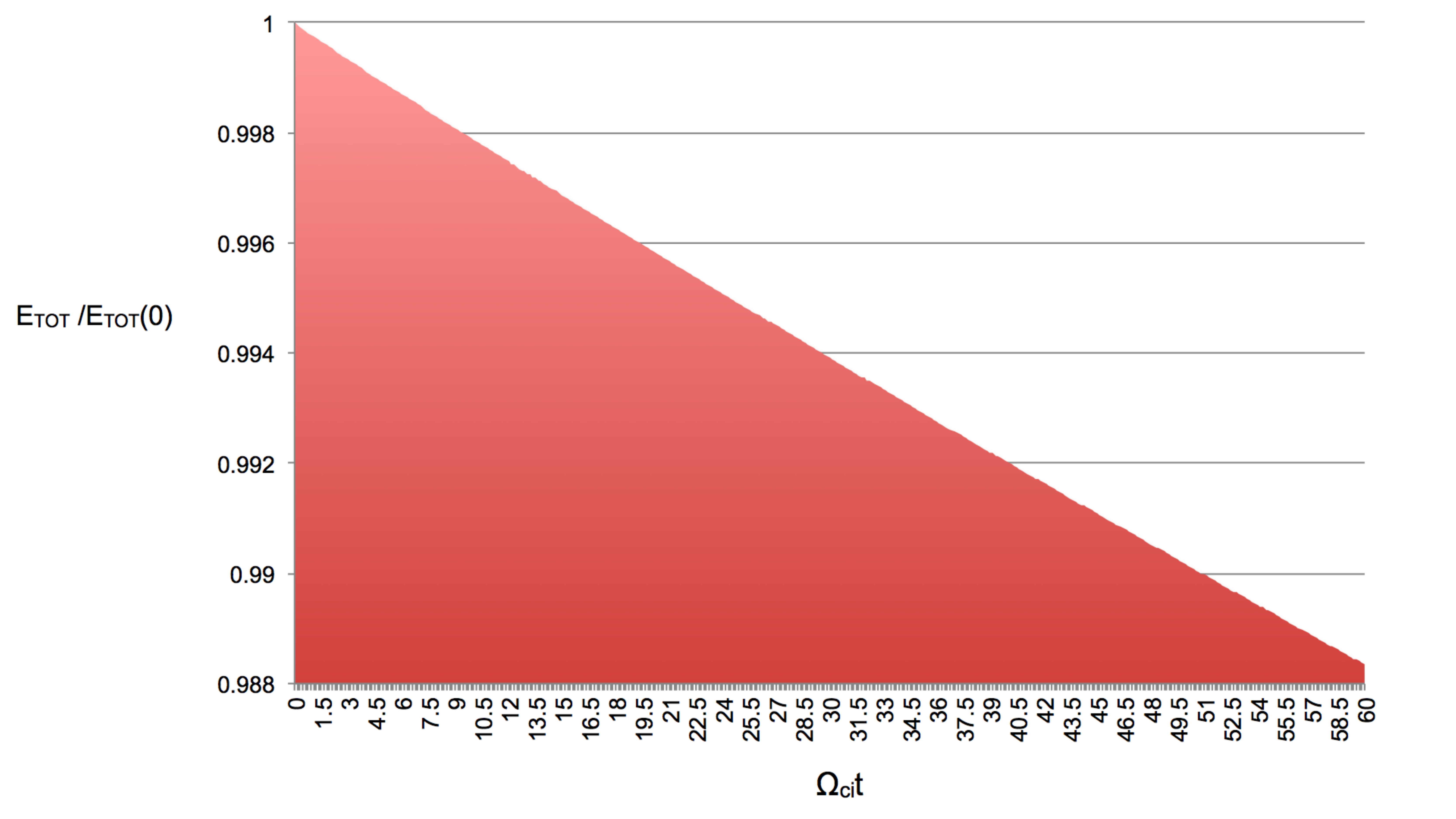}
\caption{Evolution of the total energy normalized to its initial value in the ion cyclotron frequency test. At the final time $\Omega_{ci} \ t = 60$, the total energy decreased 1.2 \%. }
\label{EnergyMaxMaxwellian}
\end{figure}

The total energy is monitored during the simulation and it is presented in Figure~\ref{EnergyMaxMaxwellian}. As predicted by the theory of implicit PIC methods, the total energy in the system decreases in time with a 1.2 \% variation at the end of the simulation. This results from the numerical damping of the unresolved waves present in the system.

\subsection{Magnetic Reconnection}
The fluid-kinetic PIC solver has been tested against the magnetic reconnection problem in two and three dimensions. Magnetic reconnection is the reorganization of the magnetic field topology with concurrent conversion of magnetic field energy into kinetic energy of plasma~\cite{biskamp2005magnetic}. It is an ubiquitous phenomenon in space, astrophysical, and laboratory plasmas. In fusion reactors, magnetic reconnection is responsible for disruptive instabilities that limit the plasma confinement needed for fusion reaction. The magnetic reconnection test is very challenging for PIC codes because this phenomenon occurs over time scale of tens of $\Omega_{ci}^{-1}$ that are much longer than plasma period. For this reason, it is important to use a time step that is not constrained by the explicit condition $\omega_p \Delta t < 2$ but it can be large enough to model magnetic reconnection.

The two-dimensional test follows the initial set-up of the GEM challenge~\cite{GEM:2001}. An Harris equilibrium is chosen as initial condition of the system. The computational particles are uniformly distributed in space and have a Maxwellian distribution in velocity with thermal velocities $0.045 \ c$ (electrons) and $0.0126 \ c$ (ions). The ion charge to mass ratio is 64 and 500 computational particles per cell are used. The grid consists of $128 \times 64$ grid cells. The simulation box is $L_x \times L_y = 20 \ d_i \times 10 \ d_i$. Periodic boundary conditions are applied in the $x$ direction, while perfect conductor for the electromagnetic field and reflecting boundary conditions are applied in the $y$ direction. The time step is $2.5 \omega_{p}^{-1}$ and de-centering parameter $\theta = 1.0$. 

\begin{figure}[ht]
\includegraphics[width=1 \columnwidth]{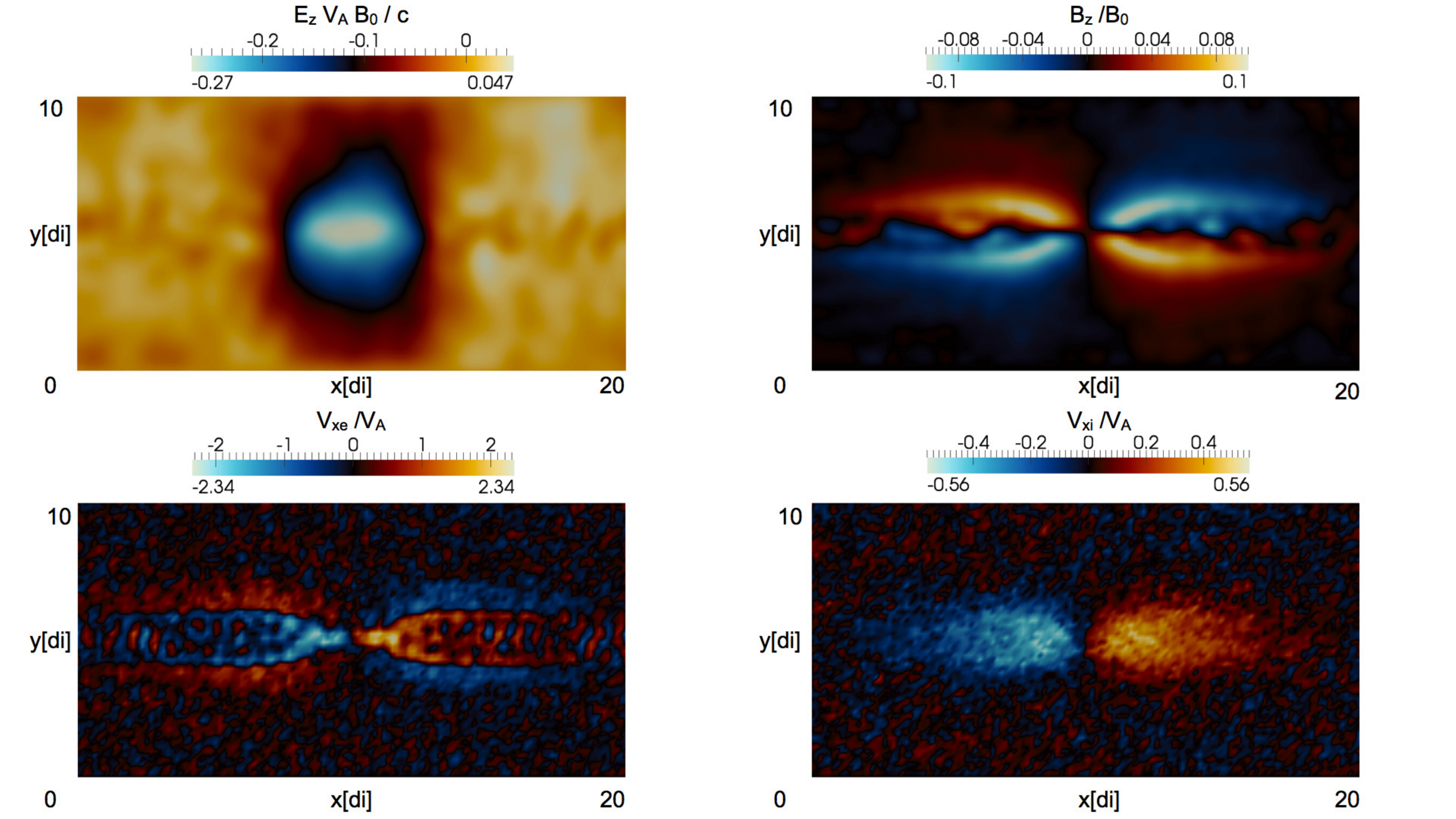}
\caption{Contour-plots of the different components of electromagnetic fields at time $\Omega_{ci} \  t = 11.7$ in the two-dimensional magnetic reconnection simulation.}
\label{recon2dcontourplot}
\end{figure}

Figure ~\ref{recon2dcontourplot} shows the different components of the electromagnetic field at time $\Omega_{ci} \  t = 11.7$. The $E_z$ component is called reconnection electric field and it is related to the speed reconnection occurs (reconnection rate). Its value at the center of the simulation box is approximately 0.3 $B_0 V_A / c$ ($V_A$ is the Alfv\'{e}n velocity) and it is an agreement with the simulation results reported in previous studies~\citep{markidis2011kinetic}. The out of plane component of the magnetic field $B_z$ shows the typical quadrupolar structure of Hall reconnection with peak values $0.1 \  B_0$. $\mathbf{V}_e$ and $\mathbf{V}_i$ are the electron and ion fluid velocities defined as $\mathbf{V}_s = \mathbf{J}_s/\rho_s$. Two electron jets exit the reconnection site in the middle of the simulation box. Their velocity reaches approximately 2.3 $V_A$. Ion jets are more diffuse and have lower velocity approximately 0.55 $V_A$. At time $\Omega_{ci} \  t = 11.7$, magnetic reconnection just started and the intensity of these jets later increases. All these results are in excellent agreement with previous simulations of the GEM challenge problem~\cite{GEM:2001}.

\begin{figure}[ht]
\includegraphics[width=1 \columnwidth]{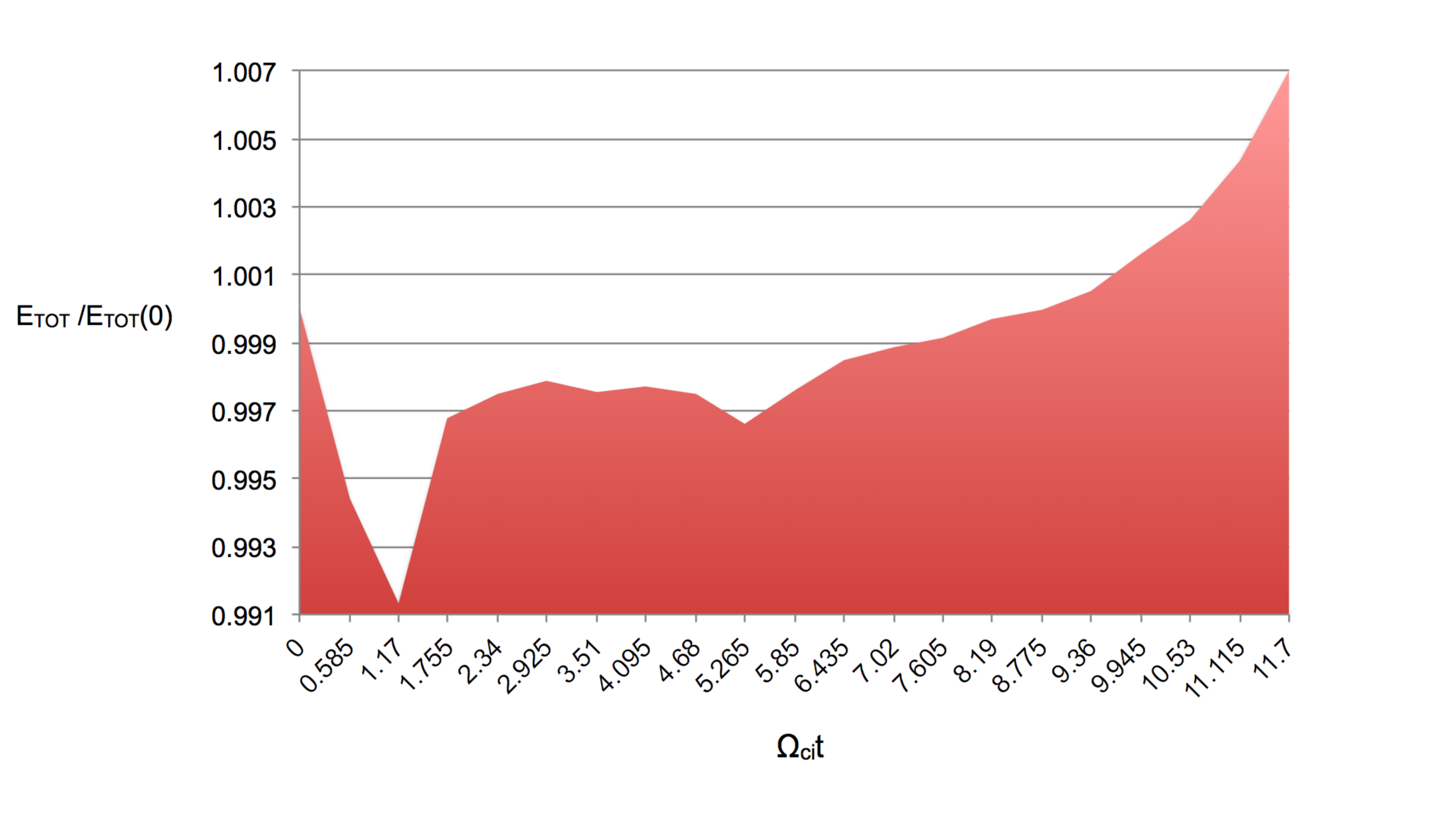}
\caption{Total energy variation in the two dimensional simulation of magnetic reconnection.}
\label{2dEnVar}
\end{figure}

The total energy of the system is rather well preserved during the simulation with a maximum variation of 0.8 \% , as shown in Figure ~\ref{2dEnVar}. The total energy initially decreases, and start increasing once magnetic reconnection develops approximately at time $\Omega_{ci} \  t \approx 9$.

\begin{figure}[ht]
\includegraphics[width=1 \columnwidth]{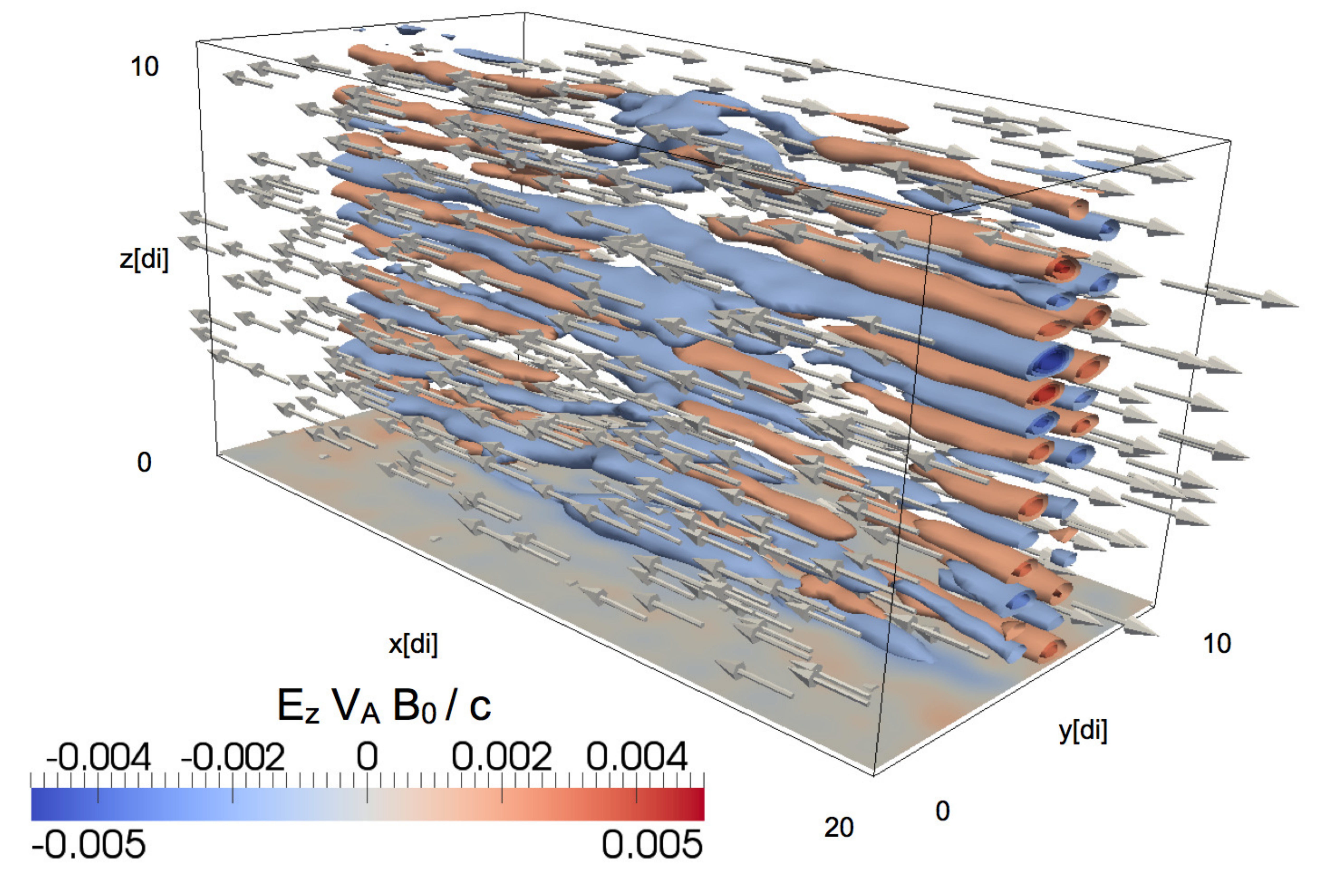}
\caption{Isosurfaces of the $z$ component of the electric field and quiver plot of the magnetic field $\mathbf{B}$ at time $\Omega_{ci} \  t = 8.77$. The plot shows features that are present only in three dimensional simulation of magnetic reconnection, such as the presence of the lower-hybrid waves propagating in the $z$ direction.}
\label{EzB15000}
\end{figure}

In addition to the two-dimensional GEM challenge, the fluid-kinetic PIC solver has been tested against magnetic reconnection in three-dimensional geometry also. The same parameters of the two dimensional simulation are used, and a third direction $z$ with $L_z=10 \  d_i$ divided in 64 cells is introduced. Periodic boundary conditions are applied in the $z$ direction. By starting from same condition, we can evaluate the effects of the third dimension on the evolution of magnetic reconnection. In particular, instabilities in the third direction arise. For instance, the presence of waves (identified as lower-hybrid waves) is clear in the isosurface plot of the $z$ component of the electric field in Figure ~\ref{EzB15000}.

\begin{figure}[ht]
\includegraphics[width=1 \columnwidth]{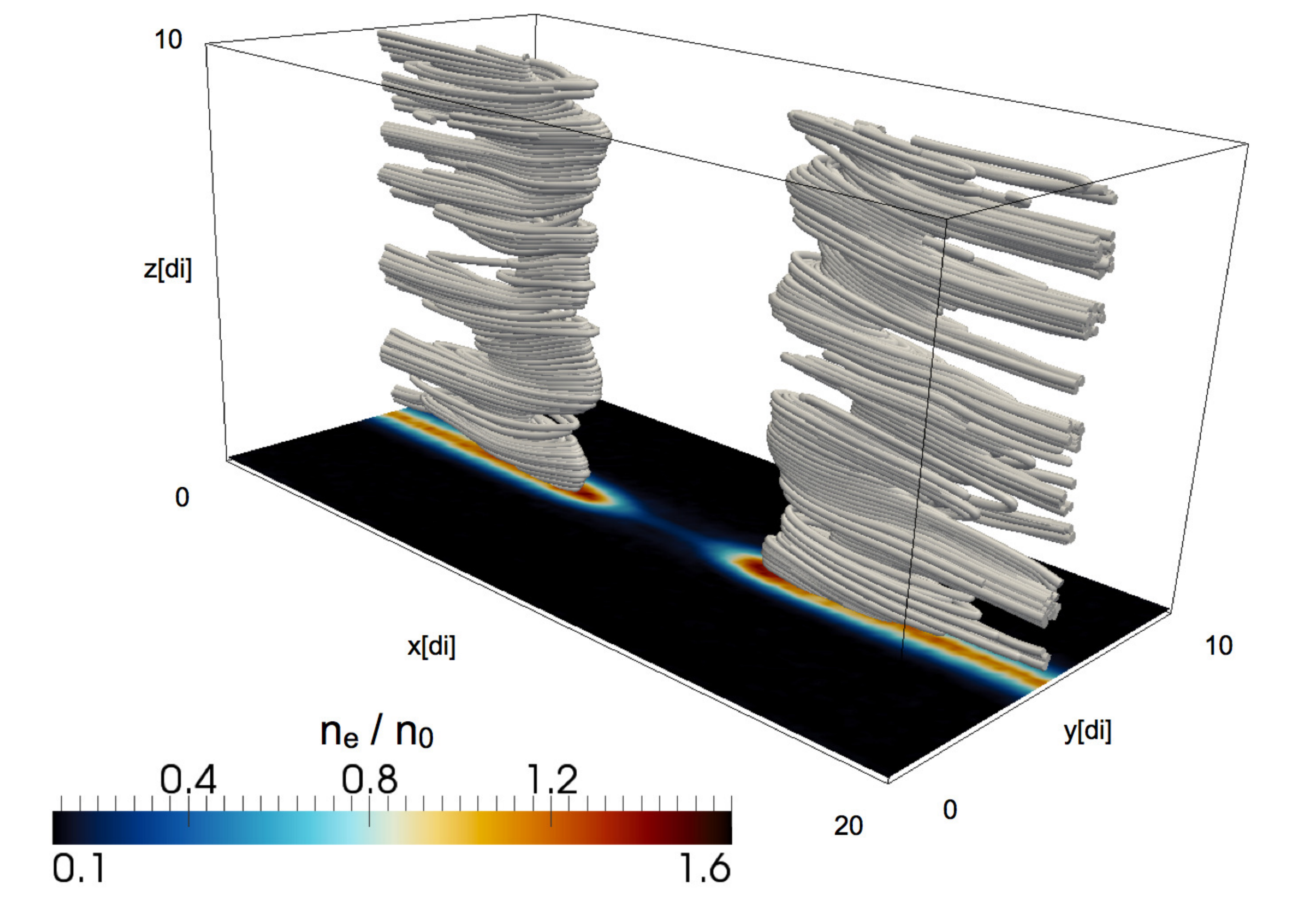}
\caption{Magnetic field lines and contour plot of the electron density $n_e$ on the plane $z = 0$ at time $\Omega_{ci} \  t = 11.7$.}
\label{Bne2000}
\end{figure}

Figure \ref{Bne2000} shows the three dimensional structure of the magnetic field represented in grey lines. A contour plot of the electron density on the plane $z=0$ is superimposed in the plot.

These three-dimensional simulation results  are in good agreement with the results of previous three-dimensional simulations~\cite{markidis2012three, vapirev2013formation}.

\section{Performance Results}
The performance of the fluid-kinetic PIC solver has been evaluated by running the two-dimensional reconnection problem presented in the previous section. The tests have been completed on a 1.7 GHz Intel Core i5 with 4 GB RAM memory running on Mac OS X 10.8.2 using the gcc 4.2 compiler. \\

To study the effect of the simulation time step on the numerical performance, three time step values were chosen: $\Delta t = 2.5 \ \omega_p^{-1}, 1.25 \ \omega_p^{-1}, 0.25 \ \omega_p^{-1}$. Typically in explicit PIC codes $\Delta t = 0.1 \ \omega_p^{-1}$, so the test with $\Delta t = 2.5 \ \omega_p^{-1}$ uses a time step that is 25 times the typical time step used in explicit PIC codes. The simulations run a different number of computational cycles in order to cover the simulation period $ t_{sim} = 62.5 \ \omega_p^{-1}$ for all the simulations. For this reason, 25 computational cycles are computed with $\Delta t = 2.5 \ \omega_p^{-1}$, 50 cycles with $\Delta t = 1.25 \ \omega_p^{-1}$ and 250 cycles with $\Delta t = 0.25 \ \omega_p^{-1}$. \\

A restarted GMRes solver is used to solve the fluid-Maxwell equations. After 20 iterations, the GMRes is restarted. The iteration stopping criterium is set by choosing a solver error tolerance $\epsilon_r$. The iterations stops when the ratio of the norm of the residual over the norm of the initial residual is less than the given solver error tolerance $\epsilon_r$. Three $\epsilon_r$ are considered in this study: $10^{-3}, 10^{-4}, 10^{-5}$. \\

Tables ~\ref{tab1} and ~\ref{tab2} summarize the performance tests for simulation with $\theta=1.0$ and $\theta=0.5$ respectively. For each time step values and solver error tolerances, the execution time in seconds for the fluid-Maxwell solver, mover and interpolation, and GMRes average number of restarts and iterations, and the variation of the total energy at last cycle in \permil   are reported.

\begin{table}[t]
\centering
\begin{tabular}{ |c | c | c  | c | c | c | c  | c | c |}
  \hline
  $\Delta t$ & $N_t$& $\epsilon_r$ & solver & mover & interp.&   rest. & it. & $\Delta E/ E_0 (\permil)$\\ \hline
  2.5	 & 25 & $10^{-3}$	& 105.8 & 204.6	& 82.5	 & 	1	& 4.14 & -1.41\\ \hline
  1.25 &	50 &	$10^{-3}$	&104.9	&405.6	&165.7	& 0		&12 	&-0.79\\ \hline
  0.25 &	125	&$10^{-3}$&	142.6	&1012.3	&413.3	&0	&	6 & -0.57 \\ \hline \hline 
  2.5	&	25	& $10^{-4}$	&149.0	&209.1&	82.9&	1	&14.24&  -1.41 \\ \hline 
  1.25&	50	&$10^{-4}$&	134.3	&411.3&	168.0	&0	&15  &  -0.79 \\ \hline 
  0.25 &	125	&$10^{-4}$&	161.9	&1017.3&	411.7&	0&	7  &-0.57\\ \hline \hline 
  2.5	& 	25	&$10^{-5}$&	233.6&	203.3&	82.3&	2	&15.08	&-1.41 \\ \hline 
 1.25	 &	50	&$10^{-5}$	&176.4&	414.1&169.7	&0	&19 &	-0.79\\ \hline 
  0.25	&	125	&$10^{-5}$&	199.7&	1016.2	&412.0	&0	&9& -0.57\\ \hline 
\end{tabular}
\caption{Performance results for a magnetic reconnection problem with $\theta=1.0$ varying the time step values and solver error tolerances. The fourth, fifth and sixth column present the execution time in seconds for the fluid-Maxwell solver, mover and interpolation stages in the PIC code. The seventh and eighth column present the average number of restarts and iterations in the GMRes solver (GMRes is restarted after 20 iterations), and the ninth column shows the final energy variation in \permil .}
\label{tab1}
\end{table}

\begin{table}[t]
\centering
\begin{tabular}{ |c | c | c  | c | c | c |  c | c | c |}
  \hline
  $\Delta t$ & $N_t$& $\epsilon_r$ & solver & mover & interp.&   rest. & it. & $\Delta E/ E_0 (\permil)$\\ \hline
  2.5	&	25	&$10^{-3}$	&63.607	&203.492&	81.3589&		0	&14.8&	-0.71 \\ \hline
  1.25&		50	&$10^{-3}$&	71.6382	&405.761	&163.599	&	0	&8.02&	-0.55  \\ \hline
  0.25&		125	&$10^{-3}$&	126.461	&1011.43	&418.004	&	0	&5 & -0.39 \\ \hline \hline
  2.5	& 25	&$10^{-4}$&	90.3208&	205.549	&83.3028	&	0	&19	& -0.57 \\ \hline
  1.25&	50&	$10^{-4}$	&86.455&	400.277&	160.966&		0	&10&  -0.55 \\ \hline
  0.25	&125	&$10^{-4}$&	141.509	&1007.13&	408.349&		0&	6 & -0.39 \\ \hline \hline
 2.5	&	25	&$10^{-5}$&	233.632&	204.848&	82.4014&		1	&10.3  & -0.59 \\ \hline
 1	&	50 &	$10^{-5}$&	89.1198&	405.792&	164.018&		0 &	10	& -0.42  \\ \hline
 0.25	 & 	125	&$10^{-5}$	&160.311&	1011.19&	411.657	&	0&	7  & -0.397 \\ \hline
\end{tabular}
\caption{Performance results for a magnetic reconnection problem with $\theta=0.5$ varying the time step values and solver error tolerances. The fourth, fifth and sixth column present the execution time in seconds for the fluid-Maxwell solver, mover and interpolation stages in the PIC code. The seventh and eighth column present the average number of restarts and iterations in the GMRes solver (GMRes is restarted after 20 iterations), and the ninth column shows the final energy variation in \permil .}
\label{tab2}
\end{table}

\begin{figure}[ht]
\includegraphics[width=1 \columnwidth]{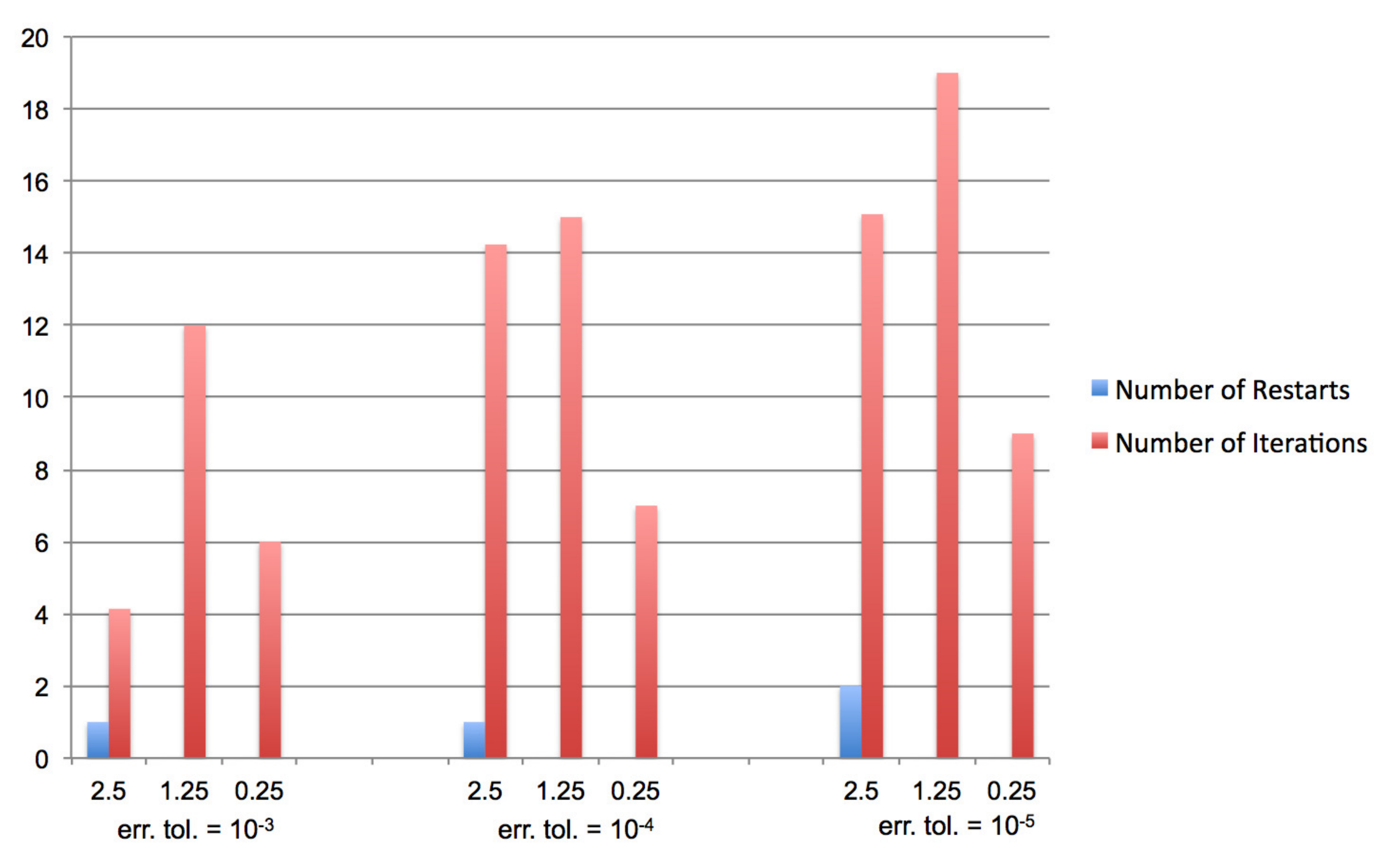}
\caption{Average number of GMRes restarts (GMRes is restarted after 20 iterations) and iterations varying the time step $\Delta t = 2.5 \ \omega_p^{-1}, 1.25 \ \omega_p^{-1}, 0.25 \ \omega_p^{-1}$, and the  solver error tolerance $\epsilon_r = 10^{-3}, 10^{-4}, 10^{-5} $ with $\theta = 1.0$. The number of GMRes restarts and iterations strongly depends on the simulation time step and on the solver error tolerance.}
\label{RestIteMF}
\end{figure}

Figure \ref{RestIteMF} shows the average number of GMRes restarts and iterations varying the time step $\Delta t = 2.5 \ \omega_p^{-1}, 1.25 \ \omega_p^{-1}, 0.25 \ \omega_p^{-1}$, and the solver error tolerances $\epsilon_r = 10^{-3}, 10^{-4}, 10^{-5} $ with $\theta = 1.0$. Analyzing the plot, it is clear that the number of solver restarts and of iterations increases when the simulation time step increases and the solver error tolerance decreases.

\begin{figure}[ht]
\includegraphics[width=1 \columnwidth]{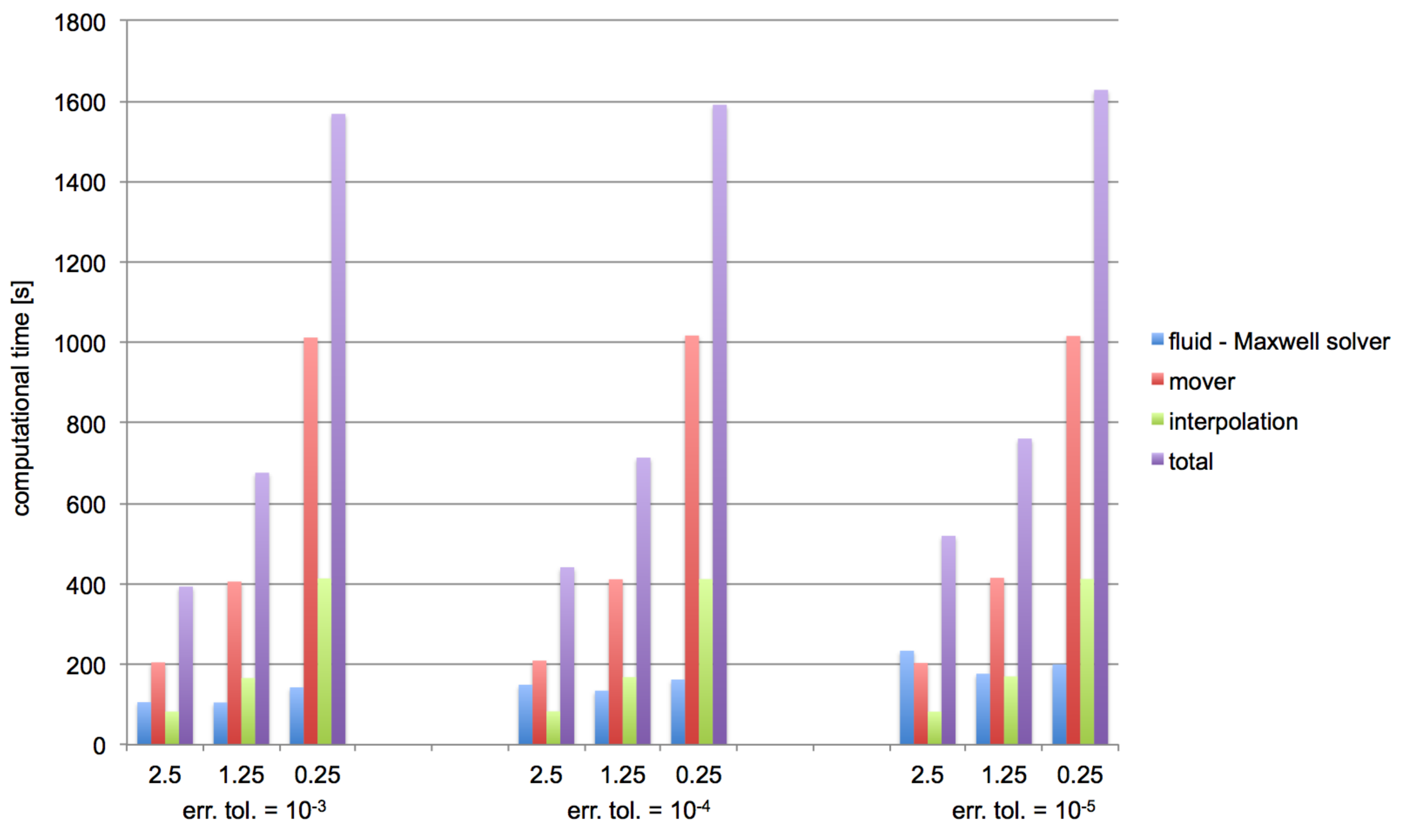}
\caption{Computational time of the different stages of the PIC code varying the time step $\Delta t = 2.5 \ \omega_p^{-1}, 1.25 \ \omega_p^{-1}, 0.25 \ \omega_p^{-1}$, and the GMRes solver error tolerance $\epsilon_r = 10^{-3}, 10^{-4}, 10^{-5} $ with $\theta = 1.0$.}
\label{CompTime}
\end{figure}

The total execution time of the different stages of the PIC code (fluid-Maxwell solver, mover, interpolation) varying the time step and solver tolerance is shown in Figure~\ref{CompTime}. As seen in Figure~\ref{RestIteMF}, the number of iterations and restarts increases when the error tolerance is decreased and the time step values is increased. Thus, accordingly the computational cost of the fluid Maxwell solver increases when the error tolerance is decreased and the time step values is increased. This is clear when analyzing Figure~\ref{CompTime}. The mover and interpolation computational costs do not depend on the time step values since a fixed number of predictor-corrector iterations is set to three for all the simulations. An analysis of the computational efficiency of the mover is presented in Ref.~\cite{markidisPHDthesis}.

\begin{figure}[ht]
\includegraphics[width=1 \columnwidth]{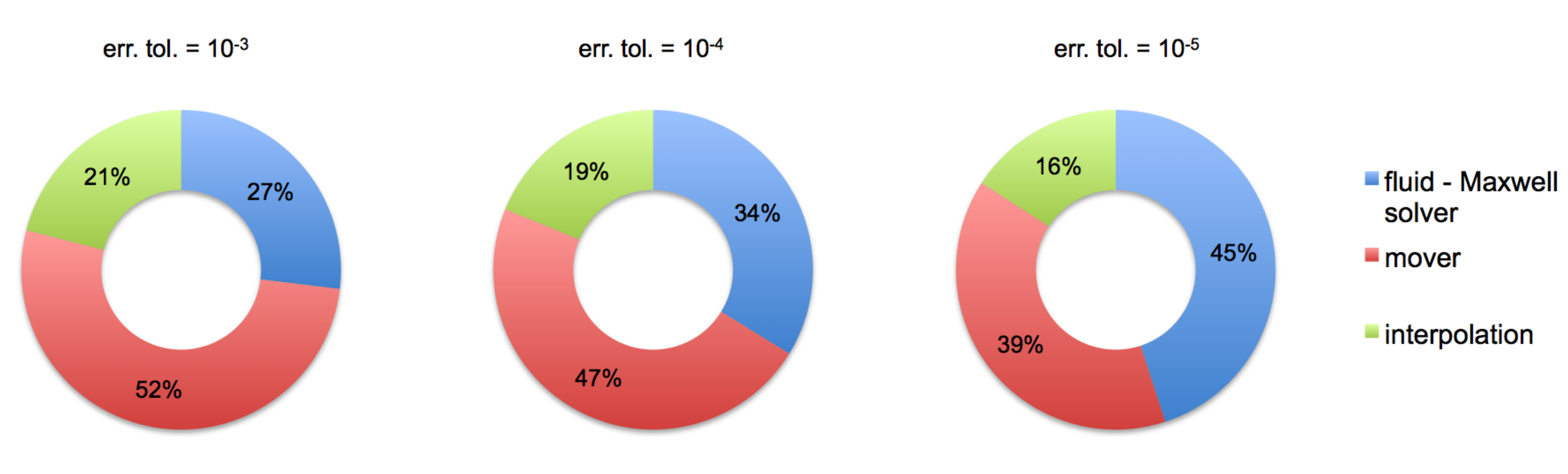}
\caption{Percentage of computational cost of the different stages of the fluid-kinetic PIC code varying the solver error tolerance $\epsilon_r = 10^{-3}, 10^{-4}, 10^{-5} $with $\Delta t = 2.5 \ \omega_p^{-1}$ and $\theta = 1.0$. The fluid-Maxwell solver requires 45\% of the total computational cost when $\epsilon_r = 10^{-5}$ is used.}
\label{CompCost}
\end{figure}

\begin{figure}[ht]
\includegraphics[width=1 \columnwidth]{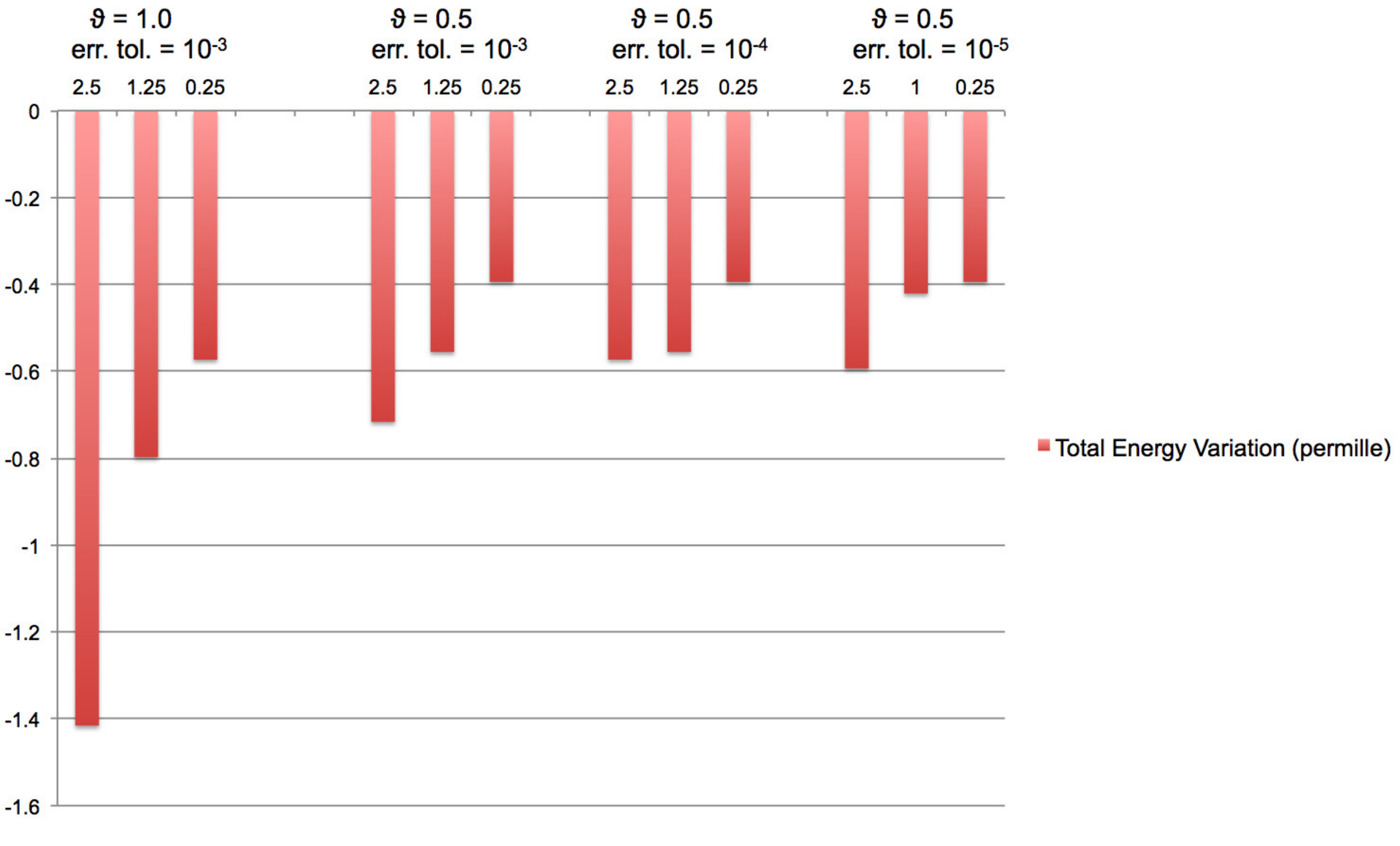}
\caption{Total energy variation in $\permil$ varying the time step $\Delta t = 2.5 \ \omega_p^{-1}, 1.25 \ \omega_p^{-1}, 0.25 \ \omega_p^{-1}$, and the GMRes solver error tolerance $\epsilon_r = 10^{-3}, 10^{-4}, 10^{-5} $ with $\theta = 1, 0.5$. The fluid-kinetic PIC solver is numerically dissipative. Total energy conservation depends on the time step and on the solver error tolerance.}
\label{EnVarPerf}
\end{figure}

As in other semi-implicit methods, the solver for computing the electromagnetic fields constitutes a not negligible part of whole computational cost as opposed to explicit PIC codes where the Maxwell solver computation is negligible with respect to the mover computational cost. Figure~\ref{CompCost} shows the percentage of computational cost of the different stages of the fluid-kinetic PIC solver varying the solver error tolerance with $\theta = 1.0$.
Figure \ref{EnVarPerf} shows the final variation of the total energy in $\permil$ normalized to the initial total energy. It is clear from Figure~\ref{EnVarPerf} that the numerical damping is stronger in the case of $\theta=1.0$ and decreases by reducing the time step as predicted by the numerical dispersion relation~\cite{brackbill-cohen-85}. \\

Finally, the performance of the fluid-Maxwell solver has been compared to the performance of the implicit moment Maxwell solver used in the iPIC3D code~\cite{Ricci2002,iPIC} for the test problems presented in Section 6. It has been found that the two GMRes solvers converge with approximately the same number iterations given the same solver error tolerance. However, because the number of unknowns and operations is larger in the fluid-Maxwell solver, the Maxwell solver of the implicit moment method results faster than the fluid-Maxwell solver. As future work, the equations of Maxwell solvers and the equation to estimate current and charge density in the implicit moment method will be used to precondition the fluid-Maxwell solver in the fluid-kinetic PIC method.

\section{Discussion and Conclusions}
A fluid solver that retains kinetic effects by calculating the stress tensor with computational particles has been presented in the context of future coupled fluid-kinetic plasma simulations. The fluid continuity and momentum equations together with the second order formulation of Maxwell's equations are solved concurrently using the finite difference box scheme.  \\

The proposed scheme is a promising approach for coupling fluid and kinetic methods in a unified framework. This work shows that one of the main challenges in coupling kinetic and fluid approaches is to make the charge and current densities from fluid and kinetic models consistent when both descriptions are present. This problem arise in the interface regions between coupled fluid-kinetic simulations~\cite{SWIFF:2013}. In fact, difference between densities calculated by solving the fluid equations and by particle interpolation originate. We showed that by taking the densities from particles before the fluid-Maxwell solver or by enforcing densities from fluid solver and from particle interpolation to be the same at a given computational cycle, the PIC simulation remains stable. An alternative method is proposed in Refs.~\cite{Vedin:2006,Vedin:2007} and consists of correcting the electric field acting on the particles to ensure the equality of the fluid and kinetic densities. \\

Overall, we have identified four main advantages of the fluid-kinetic PIC solver:
\begin{enumerate}
\item It includes both fluid and kinetic models in the same framework making the fluid-kinetic PIC solver an ideal candidate for coupling fluid-kinetic models.
\item It shows that fluid-Maxwell solvers for plasma simulations that neglect kinetic effects can be equipped with computational particles to introduce kinetic effects in the simulation. Existing multi-fluid codes can be easily extended using particles to calculate the pressure tensor instead of using ad-hoc closure equations. 
\item Because of the implicit discretization in time, the fluid-kinetic PIC solver allows to use large time steps enabling kinetic simulations over time scales typical of the fluid simulations. In the result section it has been shown that using time steps largely exceeding the values used in typical explicit PIC codes still leads to numerical stability and to the correct results.
\item It simplifies the formulation of the semi-implicit PIC models making the implementation of such schemes less prone to errors. The wide-spread use of semi-implicit PIC methods has been limited by the complexity of their software development. In addition, the fluid-kinetic PIC model removes an approximation of moment implicit methods by including the direct response of charge and current densities to a change of the electric field directly in the solver iteration.
\end{enumerate}

Finally, we note that the fluid-kinetic PIC solver can be further improved by calculating the stress tensor $\mathcal{T}$ in Eq.~\ref{FluidDiscretized} by evaluating it at time level $n+1/2$ instead of time level $n$. This can be achieved by using the kinetic enslavement technique~\cite{taitanoMSthesis,Markidis:2011}. In this method at each solver iteration, the estimate of the electric and magnetic fields can be used to calculate an estimate of particle positions and velocities at time level $n+1/2$ and therefore of the stress tensor at time level $n+1/2$. However, at each solver iteration particle equations of motion are solved requiring a very high computational effort. In addition, the kinetic enslavement technique needs the use of a non linear solver because the interpolation step directly in the solver iteration introduces a non linearity. It is not clear that the computational effort required by kinetic enslavement will result in an effective improvement of the method. The study of the numerical advantages of the kinetic enslavement in the proposed fluid-kinetic PIC method will be a topic of future research.

\section*{Acknowledgment}
The present work is supported by the European CommissionÕs Seventh Framework Programme (FP7/2007-2013) under the grant agreements no.~287703 (CRESTA, cresta-project.eu) and no.~263340 (SWIFF project, www.swiff.eu). S.M. acknowledges support from LUTH at Observatory of Paris.

\bibliographystyle{elsarticle-num}

\providecommand{\noopsort}[1]{}\providecommand{\singleletter}[1]{#1}%

\end{document}